\documentclass[preprint, amsmath,amssymb,nofootinbib]{revtex4}
\usepackage{graphicx}
\usepackage{array}

\begin{document}

\title{Footprints of Statistical Anisotropies} 
\author{C. Armendariz-Picon}
\email{armen@phy.syr.edu}
\affiliation{Physics Department, Syracuse University. \\ \today}

\begin{abstract}
We propose and develop a formalism to describe and constrain statistically anisotropic primordial perturbations. Starting from a decomposition of the primordial power spectrum in spherical harmonics,   we find how the temperature fluctuations observed in the CMB sky are directly related to  the coefficients in this harmonic expansion. Although the angular power spectrum does not discriminate between statistically isotropic and anisotropic perturbations, it is possible to define analogous quadratic estimators that are direct measures of  statistical anisotropy. As a simple illustration of our formalism we  test for  the existence of a preferred direction in the primordial perturbations using full-sky CMB maps. We do not  find  significant evidence supporting the existence of a dipole component in the primordial spectrum.

\end{abstract} 

\maketitle

\section{Introduction}

At the time Einstein decided to apply his equations to the universe, observational data in cosmology was rather scarce, if not inexistent. Hence, instead of  relying on observations to constrain the spacetime metric, he postulated the ``cosmological principle," the isotropy and homogeneity of the universe. Similarly, guided by theoretical prejudice, Einstein assumed that the universe was static, which, incidentally, is what forced him to introduce a cosmological constant in his equations. It turns out that the universe is actually homogeneous and isotropic on large scales, as confirmed, for  example, by the homogeneity of the distribution of luminous red galaxies \cite{homogeneity} and the isotropy of the cosmic microwave background \cite{WMAP}. However, the universe is not static, and Einstein's staticity assumption prevented him from predicting the expansion of the universe later discovered by Hubble.

The cosmological principle has a similar, but formally independent counterpart for cosmological perturbations. It states that perturbations are statistically homogeneous and isotropic, that is, that their correlation functions are invariant under translations and rotations. These assumptions are so  ingrained and integrated into our treatment of cosmological perturbations, that often we are unaware of them. As a consequence, the statistical homogeneity and isotropy of cosmological perturbations has received little attention in the literature, and only recently the large scale anomalies observed in the cosmic microwave background radiation (CMB) have led to a spurt of interest in statistical anisotropy.  

In fact, the data  provided by  the WMAP  cosmic microwave  background experiment \cite{WMAP}, shows hints of violations of statistical isotropy. These hints include  the alignment of  the quadrupole and octopole \cite{OTZH},  unlikely correlations between ``multipole vectors''  \cite{multipole-vectors}, a north-south  asymmetry in the CMB  sky \cite{asymmetry} and evidence for the existence of a symmetry plane  \cite{LandMagueijo, inversions}. On the other hand,  a measure of statistical anisotropy proposed in \cite{HajianSouradeep} shows no evidence for statistical anisotropy, and it has been also suggested that the previous hints are artifacts of the heavily processed maps used to analyze the data \cite{BEBGL}. 

In any case, regardless of whether the CMB sky is truly  statistically anisotropic or not, the statistical isotropy of cosmological perturbations should be subject to empirical test, rather than taken on faith. Whereas most studies have focused on the statistical isotropy of the CMB sky, in this work we analyze the isotropy of the primordial perturbations themselves.  Verifying whether cosmological perturbations are  statistically anisotropic is a way to confirm whether our basic assumptions about structure formation, and ultimately our universe, are correct. It provides a novel way to test  our current understanding of the origin of structure, embodied in inflationary models, and  it  might provide information about the universe at high energies and long wavelengths. 

In this work we  are not concerned with the mechanism responsible for the generation of statistically anisotropic primordial perturbations, although if it were impossible to seed them our analysis would be perhaps an academic exercise. The most straightforward mechanism to explain statistical anisotropies involves isotropy violations  in the background spacetime itself. These departures from isotropy  can arise from non-trivial spatial topologies \cite{topology} or  departures from the \emph{background} Friedmann-Robertson-Walker metric \cite{Bianchi}. However, these explanations require the breaking of isotropy from the onset, and they also clash with our understanding of the origin of structure. Alternatively, statistical anisotropies might arise from coherent magnetic fields in the universe \cite{magnetic}. If these magnetic fields can be regarded as perturbations in an isotropic spacetime, their contribution to the metric perturbations is quadratic in the fields, and hence presumably highly suppressed. On the other hand, if they significantly contribute to the energy density of the universe, one expects deviations from isotropy at the level of the background spacetime itself,  violating again isotropy from the onset.  In this paper we assume that the universe is homogeneous, isotropic, spatially flat and practically infinite. But even  homogeneous and isotropic universes can be dominated by sets of non-vanishing  non-scalar background fields if the latter appear in special configurations, like  for example, vector field triads.  In these universes, perturbations are expected to be statistically anisotropic \cite{triad}.  In general the existence of these non-vanishing fields signals the spontaneous breaking of Lorentz-invariance \cite{Lorentz} (rotational invariance to be more precise), so the study of statistical anisotropies in the primordial perturbations opens a  new window to test the  basic symmetries of nature. 

The paper is organized as follows. In Section \ref{sec:formalism} we precisely define what we mean by statistically isotropic or anisotropic primordial perturbations. It is in this section where we introduce a quantitative characterization of primordial statistical anisotropies. In Section \ref{sec:temperature} we connect statistical anisotropies in the primordial perturbations to statistical anisotropies in the cosmic microwave background temperature fluctuations. We define statistics that capture the amount of statistical anisotropy in the CMB and at the same time mirror the amount of anisotropy in the primordial density ripples, and we also discuss the relation between our statistics and the bipolar spectrum of Hajian and Souradeep \cite{HajianSouradeep}. Section \ref{sec:looking} contains a description of  strategies to test and constrain the statistical isotropy of the primordial perturbations. This task is simplified if there is a cut-off in the anisotropies at a particular arbitrary multipole, which is the case we mainly consider. As an illustration of the formalism, we look for the simplest departure from statistical isotropy: a preferred direction in the primordial perturbations. We apply our statistics to the full sky CMB maps and do not find evidence of statistical anisotropy. Finally, in Section \ref{sec:conclusions} we draw our conclusions. We have also included an Appendix where we summarize most of the formulae that we need in our derivations.

\section{Statistically Anisotropic Random Fields}\label{sec:formalism}

\subsection{Random Fields on Euclidean space}

Consider a  real Gaussian random field in Euclidean (flat) three-dimensional space, $\Phi(\vec{x})$.  At this point it is not important to know what $\Phi$ actually is, although the reader can think of $\Phi$  as the Newtonian potential (in longitudinal gauge \cite{MuFeBr,MaBertschinger}).  The isometries of Euclidean space consist of  translations, rotations and reflections \cite{Coxeter}. It is therefore useful to analyze how random fields behave under these transformations. The Gaussian random field  $\Phi$ is \emph{statistically homogeneous} if its momenta are invariant under translations,
\begin{equation} \label{eq:homogeneous}
	\langle \Phi(\vec{x})\rangle=\langle \Phi(\vec{x}+\vec{t})\rangle,\quad
	\langle \Phi^*(\vec{x}) \Phi(\vec{y})\rangle=\langle \Phi^*(\vec{x}+\vec{t}) \Phi(\vec{y}+\vec{t})\rangle 		\quad \forall \vec{t}\in \mathbb{R}^3,
\end{equation} 
where $\langle \, \rangle$  denotes ensemble average.\footnote{Because $\Phi$ is real by assumption, $\Phi=\Phi^*$. We include the complex conjugation in the two-point function for convenience.}  It is not essential to assume that the field is Gaussian. This assumption just allows us to concentrate on the average $\langle \Phi(\vec{x})\rangle$ and the correlation $\langle \Phi^*(\vec{x}) \Phi(\vec{y})\rangle$, which uniquely characterize the statistical properties of the random field in this case. A non-Gaussian field that does not satisfy equations (\ref{eq:homogeneous}) is not statistically homogeneous either.
 
The average $\langle \Phi(\vec{x})\rangle$ of a statistically homogeneous random field is a constant in space, and hence is trivially invariant under rotations.  A Gaussian random variable $\Phi$ is \emph{statistically isotropic} (around $\vec{O}$) if its momenta are invariant under rotations (around $\vec{O}$),
\begin{eqnarray}\label{eq:isotropic}
	\langle \Phi(\vec{x}) \rangle & = &
	\langle \Phi(\vec{O}+\mathcal{R}\cdot(\vec{x}-\vec{O})\rangle  \\ \nonumber
	\langle \Phi^*(\vec{x}) \Phi(\vec{y})\rangle &= & 
	\langle \Phi^*(\vec{O}+\mathcal{R}\cdot(\vec{x}-\vec{O})) 
	\Phi(\vec{O}+\mathcal{R}\cdot(\vec{y}-\vec{O}))\rangle \quad \forall \mathcal{R}\in SO(3).
\end{eqnarray}
If we assume that the field is statistically homogeneous, equations (\ref{eq:isotropic}) are satisfied for all points $\vec{O}\in \mathbb{R}^3$ if they are satisfied  for a single  $\vec{O}$ (say, the origin $\vec{O}=0$.) Conversely, one can also show that if a random field  $\Phi$ fulfills equations (\ref{eq:isotropic}) for all possible points $\vec{O}$, then it also fulfills equations (\ref{eq:homogeneous}). In other words, if a random field is statistically isotropic around all points, it is statistically homogeneous. Note that it is not necessary to discuss the statistical properties of $\Phi$ under inversions. The inversion of a single point can be accomplished by a translation, and the inversion of two points can be accomplished by  a rotation. 

Often it is convenient to decompose spatial functions in eigenvectors of the translation operator. In flat space these are the plane waves $\exp(i \vec{k}\cdot\vec{x})$. Let us hence consider a Fourier decomposition of the random field, 
\begin{equation}
	\Phi(\vec{x})=\int \frac{d^3k}{(2\pi)^3} \Phi(\vec{k})e^{i \vec{k}\cdot \vec{x}}.
\end{equation}
The components $\Phi(\vec{k}$) are random fields in Fourier space. In this space, the requirement of statistical homogeneity (\ref{eq:homogeneous})  translates into
\begin{equation}\label{eq:power-spectrum}
	\langle\Phi(\vec{k})\rangle=\bar{\Phi}_0 \,\delta^{(3)}(\vec{k})
	\quad \text{and} \quad
 	\langle \Phi^*(\vec{k}) \Phi(\vec{k}')\rangle= (2\pi)^3 \delta^{(3)}(\vec{k}-	\vec{k}')\frac{\pi}{2k^3}\mathcal{P}(\vec{k}),
\end{equation}
where $\bar{\Phi}_0$ is the (constant) mean of $\Phi$ and $\mathcal{P}(\vec{k})$ denotes the   power spectrum,\footnote{Note that our power spectrum has a slightly unconventional normalization.}  which is an arbitrary positive definite function\footnote{One can show this by considering the two-point function of a smoothed field $\Phi_s: 0\leq \langle\Phi^*_s(\vec{x})\Phi_s(\vec{x})\rangle$. } of $\vec{k}$. On the other hand, the requirement of statistical isotropy (\ref{eq:isotropic})   implies 
\begin{equation}
	\mathcal{P}(\mathcal{R}\vec{k})=\mathcal{P}(\vec{k}) \quad \forall \mathcal{R}\in SO(3)
\end{equation}
that is, that the power spectrum only depends on the magnitude $k$ of the wave vector $\vec{k}$. Because any positive function can be written as a square, we can express the power spectrum as 
\begin{equation}\label{eq:phi_definition}
	\mathcal{P}(\vec{k})=k^3\bar{\Phi}^*(\vec{k})\bar{\Phi}(\vec{k}),
\end{equation}
where $\bar{\Phi}$ is a real function by definition. Then, the random variable $\Phi$ is statistically isotropic if and only if $\bar{\Phi}$ only depends on $k$.  

In this paper, we consider statistically anisotropic primordial perturbations, that is, primordial perturbations whose power spectrum does not only depend on the magnitude $k$. It is hence going to be useful to decompose $\bar{\Phi}(\vec{k})$  into components that transform differently under rotations of the wave vector, 
\begin{equation}\label{eq:phi_expansion}
	\bar{\Phi}(\vec{k})=\sqrt{4\pi}\sum_{lm}\bar{\Phi}_{lm}(k) Y_{lm}(\hat{k}),
\end{equation} 
where $\hat{k}=\vec{k}/k$ and the $Y_{lm}$ are spherical harmonics.  Because $\bar{\Phi}(\vec{k})$ is real, the complex functions $\bar{\Phi}_{lm}$ satisfy $\bar{\Phi}_{lm}^*=(-1)^{m}\bar{\Phi}_{l-m}$. This decomposition of $\bar{\Phi}$  carries over to the power spectrum itself. Inserting equation (\ref{eq:phi_expansion}) into equation  (\ref{eq:phi_definition}) and  using relation (\ref{eq:three_spherical}) we find that  
\begin{eqnarray}
\mathcal{P}(\vec{k}) &=& \sqrt{4\pi}\sum_{lm} \mathcal{P}_{lm}(k)Y_{lm}(\hat{k}), \quad \text{where}
	\label{eq:spherical_power}\\
	\mathcal{P}_{lm}(k) & =& \sum_{l_1m_1,l_2m_2} (-1)^{m_1}
	k^3\,\bar{\Phi}^*_{l_1m_1}(k)\bar{\Phi}_{l_2m_2}(k)
	D(l_1 {}-m_1; l_2, m_2 | l, m), \label{eq:P_lm}
\end{eqnarray} 
and the real coefficient $D$ is given by equation (\ref{eq:D}). If the random  field is statistically isotropic, $\mathcal{P}_{lm}=0$ for $l\geq 1$. On the contrary, if $\Phi$ is statistically anisotropic, there exists at least one $l\geq 1$ and $m\in \{-l,\ldots, l\}$  such that $\mathcal{P}_{lm}\neq 0$. For simplicity we shall express our results mostly in terms of $\mathcal{P}_{lm}$, though the reader should be aware that the actual free parameters that characterize the statistical anisotropies are the $\bar{\Phi}_{lm}(k)$. Observe that because $\mathcal{P}$ is real,  $\mathcal{P}_{lm}^*=(-1)^m \mathcal{P}_{l -m}$, which also follows from  the relation (\ref{eq:P_lm}).

\subsection{Random Fields on a Sphere}

Sometimes, one is interested in random fields $\Delta(\hat{n})$ that are defined not on $\mathbb{R}^3$, but rather on a two-sphere, $\hat{n}\in \mathbb{S}^2$ . The standard  case is the temperature fluctuations on the cosmic microwave background sky. Because the isometries of a sphere consist only of rotations and inversions, for Gaussian fields it suffices to consider the properties of the random field under rotations (the behavior of the CMB under inversions has been studied in \cite{inversions}.) A Gaussian random field is \emph{statistically isotropic} if the average $\langle \Delta(\hat{n})\rangle$ and the two-point function $\langle \Delta^*(\hat{n})\Delta(\hat{m})\rangle$ are invariant under rotations,
\begin{equation}\label{eq:sphere_isotropic} 
	\langle \Delta(\hat{n})\rangle=\langle \Delta(\mathcal{R}\cdot\hat{n})\rangle,\quad
	\langle \Delta^*(\hat{n}) \Delta(\hat{m})\rangle=\langle \Delta^*(\mathcal{R}\cdot\hat{n}) 	\Delta(\mathcal{R}\cdot\vec{m})\rangle \quad \forall \mathcal{R}\in SO(3).
\end{equation}
In the applications we have in mind, random fields on a sphere receive weighted contributions from an infinite number of Fourier modes. For instance,  the temperature fluctuations measured by an observer at $\vec{x}_0$ in the direction $\hat{n}$ can be written as \cite{MaBertschinger}
\begin{equation}\label{eq:delta_expansion}
	\Delta_T(\vec{x}_0,\hat{n})=\int \frac{d^3k}{(2\pi)^3} \Delta(\vec{k},\hat{n}) 		\Phi(\vec{k})e^{i\vec{k}\cdot\vec{x}_0},
\end{equation}
where $\Phi$ is a random field in Fourier space and $\Delta(\vec{k},\hat{n})$ is a real function. We shall specify what $\Phi$ and $\Delta(\vec{k},\hat{n})$ represent below. At this point, we just want to know what constraints  statistically isotropy of  $\Delta_T(\vec{x}_0,\hat{n})$ imposes on the quantities $\Delta(\vec{k},\hat{n})$. First, note that it follows from equation (\ref{eq:delta_expansion}) that if $\Phi$ is statistically homogeneous,  
$\langle \Delta_T(\vec{x}_0,\hat{n})\rangle$ is invariant under rotations if $\Delta(0,\mathcal{R}\hat{n})=\Delta(0,\hat{n})$. Furthermore, inserting the expansion (\ref{eq:delta_expansion}) into the second equation in  (\ref{eq:sphere_isotropic}) we arrive at the condition
\begin{equation}
	\int \frac{d^3k}{(2\pi)^3}\left[
	|\Delta(\vec{k},\hat{m})|^2\mathcal{P}(\vec{k})-
	|\Delta(\mathcal{R}\vec{k},\mathcal{R}\hat{n})|^2\mathcal{P}(\mathcal{R}\vec{k})
	\right]=0.
\end{equation}
Therefore, $\Delta_T(\vec{x}_0,\hat{n})$ is statistically isotropic if
\begin{equation} \label{eq:origins}
	\Delta(\mathcal{R}\vec{k},\mathcal{R}\hat{n})=
	\Delta(\vec{k},\hat{n})
	\quad \text{and} \text \quad
	\mathcal{P}(\mathcal{R}\hat{k})=\mathcal{P}(\vec{k}) 
	\quad \forall \mathcal{R}\in SO(3).
\end{equation}
Note that the two-point function $\langle \Delta_T^*(x_0,\hat{n}) \Delta_T(x_0,\hat{m})\rangle$ does not depend on $\vec{x}_0$ if the random field $\Phi$ is statistically homogeneous. 

\section{Temperature Anisotropies}\label{sec:temperature}

Our main concern here are the contribution of scalar perturbations to the temperature fluctuations in the cosmic microwave background. Let us denote by $\Delta_T(\vec{x},\hat{n})$  the temperature fluctuations  measured by an observer at $\vec{x}$ looking at direction $\hat{n}$ in the sky.  As hinted above, it is more convenient to study the Fourier transform of the previous fluctuations, $\Delta_T(\vec{k},\hat{n})$. In linear perturbation theory, the evolution of $\Delta_T(\vec{k},\hat{n})$ is described by a set of decoupled differential equations for each mode $\vec{k}$. The linearity of these equations allows us to separate the initial conditions from the evolution. In particular, for adiabatic initial conditions one can write
\begin{equation}
	\Delta_T(\vec{k},\hat{n})=\Delta(\vec{k},\hat{n}) \Phi(\vec{k}),
\end{equation}
where $\Delta(\vec{k},\hat{n})$ is the temperature fluctuation one obtains by evolving the perturbation  equations from initial conditions where $\Phi(\vec{k})=1$, and $\Phi(\vec{k})$ is the value of the Newtonian potential at a sufficiently early time, well into the radiation dominated era. Note that this split between evolution and initial conditions is arbitrary to some extent: what we consider as initial conditions here might be regarded for instance as the  outcome of an inflationary stage. 

Recall now our discussion of statistical isotropy of a random field defined on a sphere. It follows from equation (\ref{eq:origins})  that  violations of statistical isotropy in the CMB sky might originate from two sources: from the initial conditions (encoded in the power spectrum $\mathcal{P}$), or from the evolution (encoded in the transfer function $\Delta(\vec{k},\hat{n})$). In this work we assume that the evolution does not generate statistically anisotropic perturbations, that is, we assume that $\Delta(\vec{k},\hat{n})=\Delta(\mathcal{R}\vec{k},\mathcal{R}\hat{n})$ for all rotation matrices $\mathcal{R}$ in $SO(3)$. This is the case in cosmological models where the universe does not contain any non-scalar background field, like in the standard $\Lambda$CDM scenario.\footnote{In cosmological scenarios that contain non-scalar background  fields, it is possible to construct scalar perturbations that are not invariant under rotations of $\vec{k}$ \cite{triad}. For example, if there exists a background vector field $\vec{A}$ in the universe, the scalar  $\vec{k}\cdot\vec{A} \, \Phi$ is not invariant under $\vec{k}\to \mathcal{R}\cdot\vec{k}$.} As a consequence,   the differential equations that describe the evolution of the perturbations  are invariant under rotations of $\vec{k}$ and $\hat{n}$, and it follows that  $\Delta_T(\vec{k},\hat{n}$) is a function of the two scalars one can construct out of $\vec{k}$ and $\hat{n}$, namely $k$ and $\hat{k}\cdot\hat{n}$, $\Delta(\vec{k},\hat{n})=\Delta(k,\hat{k}\cdot\hat{n})$. This property then allows us to expand $\Delta$ in Legendre polynomials $P_l$,
\begin{equation}\label{eq:expansion_Legendre}
	\Delta(\vec{k},\hat{n})=
	\sum_l(2l+1)(-i)^l\Delta_l(k)P_l(\hat{k}\cdot \hat{n}).
\end{equation}
This expansion will turn to be useful to compute the two-point function of the temperature anisotropies.

\subsection{Angular Two-point Function}
It is customary to characterize the properties of the temperature fluctuations by the angular two-point function $\langle a^*_{lm} a_{l'm'}\rangle$,  where the $a_{lm}$ are the coefficients in the spherical harmonic decomposition
\begin{equation}
	\Delta_T(\vec{x}_0,\hat{n})=\sum_{lm}a_{lm} Y_{lm}(\hat{n}),
\end{equation}
and $\vec{x}_0$ is our position in space (because of statistical homogeneity, the statistical properties of the temperature fluctuations do not depend on $\vec{x}_0$.) We would like to express this angular two-point function in terms of the initial power spectrum. Using the expansion (\ref{eq:expansion_Legendre}) and the addition theorem (\ref{eq:addition-theorem}) we find first 
\begin{equation}
	a_{lm}=4\pi (-i)^l \int\frac{d^3k}{(2\pi)^3} \Delta_{l}(k) \Phi(\vec{k})Y^*_{lm}(\hat{k}).
\end{equation}
Then,  using  the definition of the power spectrum (\ref{eq:power-spectrum}) and its expansion in spherical harmonics (\ref{eq:spherical_power})  we arrive at our sought result
\begin{eqnarray}
	\langle a^*_{l_1m_1}a_{l_2 m_2}\rangle &=& (-i)^{l_2-l_1}
	\sum_{lm}
	D(l_1, m_1; l, m | l_2, m_2) K(l_1,l_2; l,m), \quad \text{where}
	\label{eq:two-point} \\
	K(l_1, l_2; l, m) &=&	
	\int \frac{dk}{k}\Delta^*_{l_1}(k)\Delta_{l_2}(k)
	\mathcal{P}_{lm}(k), \label{eq:K}
\end{eqnarray}
and we have used  equation (\ref{eq:three_spherical}) in the Appendix.  The structure of the two-point function can be understood in the language of addition of angular momenta. The $l, m$ multipole of the primordial anisotropies can be thought of as a state with angular momentum quantum numbers $l$ and $m$.  The temperature perturbation $a_{l_1 m_1}$ ``creates"  angular momentum with quantum numbers $l_1,m_1$, and, similarly,  $a^*_{l_2 m_2}$ destroys angular momentum with quantum numbers $l_2, m_2$. Therefore, the contribution of $\mathcal{P}_{lm}$ to the two-point function  vanishes if the angular momenta $l_1, m_1$ and $l, m$ cannot add to $l_2, m_2$. In fact,  $D$ is zero unless $m_2=m_1+m$ and $l_2\in\{ |l_1-l |,\ldots, l_1+l\}$.

\subsection{Large Scale Anisotropies}

In general it is only possible to compute the  transfer functions $\Delta_{l}$ in equation (\ref{eq:K}) numerically.  In order to provide a better handle on statistical anisotropies,  in the following we consider a limit where it is possible to analyze them analytically, namely, on large angular scales. As it is well known, on these scales, neglecting the integrated Sachs-Wolfe effect, the Sachs-Wolfe relation
\begin{equation}\label{eq:delta_approximation}
	\Delta_l(k)\simeq\frac{j_l(kD_s)}{3}
\end{equation}	
 holds \cite{MaBertschinger}, where $D_s$ denotes the  comoving distance to last scattering.\footnote{By ``last scattering" we mean the decoupling of photons and matter that occurred around redshift ${z=1100}$. } 
It is important to realize that this result is an approximation with a limited range of validity. In particular, at $l<8$ the integrated Sachs-Wolfe effect  modifies\footnote{We assume a non-zero cosmological constant with $\Omega_{\Lambda}\approx 0.7$. Since we do not know what is causing cosmic acceleration, this is just a guess. Other forms of dark energy lead to different ISW effects.} the relation (\ref{eq:delta_approximation}) \cite{KofmanStarobinsky}, and at $l>30$, the baryon velocity significantly contributes to the temperature anisotropies \cite{HuSugiyama}. Therefore, the  approximation  (\ref{eq:delta_approximation}) is only valid around $8\leq l \leq 30$.

Just as one parametrizes the primordial spectrum in terms of an amplitude and a spectral index, let us parameterize the different components of the anisotropies  $\bar{\Phi}_{lm}$ as power laws. It follows then from equation (\ref{eq:P_lm}) that $\mathcal{P}_{lm}$ is a superposition of different power-laws, with different amplitudes and spectral indices. Let us assume for simplicity, and without loss of generality that
\begin{equation}
	\mathcal{P}_{lm}=A_{lm} \left(\frac{k}{k_*}\right)^{n-1}.
\end{equation}
In this case, the integral in equation (\ref{eq:K}) can be explicitly evaluated. Using equations (11.4.33) and (15.1.20) in \cite{AbramowitzStegun} we arrive at
\begin{equation}\label{eq:explicit_K}
	K(l_1, l_2; l m)=
	\frac{2^{n-4}\,\pi\,\Gamma[3-n]\Gamma\left[(l_2+l_1+n-1)/2\right]}
	{9\,\Gamma\left[(l_1-l_2+4-n)/2\right]
	\Gamma\left[(l_2-l_1+4-n)/2\right]
	\Gamma\left[(l_2+l_1+5-n)/2)\right]}
	\frac{A_{lm}}{(k_* D_s)^{n-1}}.
\end{equation}
Hence, given the amplitude $A$ and the spectral index in any given anisotropy multipole, one can exactly predict what the two-point function on large angular scales should be. In general however, we do not know what $A$ or $n$, are. Under these circumstances, equation (\ref{eq:explicit_K}) in combination with (\ref{eq:two-point}) could be used to fit for the values of $A$ and $n$ (see below.)

\subsection{Angular Power-Spectrum}
If primordial perturbations are statistically isotropic ($\mathcal{P}_{lm}=0$ for $l\geq 1$), one can easily verify that the two point function (\ref{eq:two-point}) is diagonal.  In this case it is customary to define the angular power spectrum $C_l$ through
\begin{equation}\label{eq:conventional_C}
	\langle a^*_{l_1 m_1} a_{l_2 m_2}\rangle \equiv C_{l_1} \, \delta_{l_1l_2}\delta_{m_1 m_2}.
\end{equation}
But if perturbations are statistically anisotropic the two-point function is not diagonal, and the  definition of the angular power spectrum above is meaningless. In any case, one never measures the $C_l$'s directly. Instead, one uses an appropriate estimator to infer their values. Here we shall define the $C_l$'s as the expectation value of  the estimator that is commonly used to compute the latter,
\begin{equation}\label{eq:C_l}
	C_l\equiv \langle \hat{C}_l\rangle \equiv  \frac{1}{2l+1}\sum_{m=-l}^{l} \langle a^*_{lm}a_{lm}\rangle. 
\end{equation}
The angular power spectrum defined this way is invariant under rotations.  It reduces to (\ref{eq:conventional_C}) if perturbations are statistically isotropic and it  is also closer to the quantities that we actually measure.  Note however that in the presence of statistical anisotropies, the $C_l$'s defined in this way  do not completely characterize  the statistical properties of the temperature fluctuations, since the two-point function is not diagonal. 

We want to find out now how the angular power spectrum is related to the primordial power spectrum. To this end, we substitute  equation (\ref{eq:two-point}) into the definition (\ref{eq:C_l}). Using then the identity equation (\ref{eq:D_identity}) we get
\begin{equation}
	C_l=\int \frac{dk}{k} \,|\Delta_l(k)|^2 \, \mathcal{P}_{00}(k).
\end{equation}
Therefore, $\mathcal{P}_{00}$ acts like an effective power spectrum.  If primordial perturbations are statistically isotropic, then the effective power spectrum reduces to the conventional power spectrum, $\mathcal{P}_{00}(k)=\mathcal{P}(k)$.
Otherwise, the effective power spectrum receives contributions from all anisotropy components,
\begin{equation}
	\mathcal{P}_{00}=\sum_{l,m} k^3 \,\bar{\Phi}_{l m}^* \bar{\Phi}_{l m},
\end{equation}
where we have used equations (\ref{eq:orthonormality}) and (\ref{eq:complex-conjugation}). Let us stress that the angular power spectrum defined in equation (\ref{eq:C_l}) does not differentiate between the different anisotropy multipoles. A scale invariant monopole perturbation ($\bar{\Phi}_{00}$) leads to the same set of $C_l$'s as a scale invariant anisotropy in any other multipole ($\bar{\Phi}_{lm}$). Hence, conventional numerical codes, such as CMBFAST \cite{SeljakZaldarriaga}, or CMBEASY \cite{Doran} can be used to determine the imprints of statistical anisotropies on the angular power spectrum. Alternatively, one can use equation (\ref{eq:explicit_K}) to explicitly evaluate the $C_l$'s on large---but not too large---angular scales.

\section{Looking for Primordial Statistical Anisotropies}\label{sec:looking}
\renewcommand\arraystretch{0.6}

The power in the different multipoles of the primordial perturbations $\mathcal{P}_{lm}(k)$ uniquely  determines the different two-point functions of the temperature anisotropies. The latter are observables, provided  that sufficient realizations of the random field are available.\footnote{Since we have access to only one universe, this condition is not automatically guaranteed. See the discussion of cosmic variance below.} Therefore, one can hope to determine  what the $\mathcal{P}_{lm}$ are, or at least, place upper limits on their amplitude. This section is devoted to the ultimate goal of finding ways to measure $\mathcal{P}_{lm}$. 

\subsection{Determining K}
The first obstacle we have to find in our quest to measure $\mathcal{P}_{lm}$ is that the power is hidden behind a convolution,  equation (\ref{eq:K}). Even if primordial perturbations are statistically isotropic it is difficult to invert that integral in order to find the primordial power \cite{HuOkamoto}. Therefore, as a first step towards our goal we shall concentrate in determining the integral $K$ in equation (\ref{eq:K}). Because $K(l_1, l_2; l, m)$ vanishes when $\mathcal{P}_{lm}=0$, we can think of $K$ as a measure of the anisotropic power;  a non-vanishing $K$ for $l\neq 0$ would automatically imply the existence of statistically anisotropic primordial perturbations. The advantage of focusing on $K$ is that we do not have to know what the transfer functions $\Delta_l(k)$ are. Therefore we can analyze all angular scales without having to make any assumptions about the underlying cosmology. 

The task of determining some of the values of $K$  is greatly simplified when the anisotropies have a cut-off at at multipole $l=L$, $\bar{\Phi}_{lm}=0$ for $l>L$. In this case, it follows from equation (\ref{eq:spherical_power}) that the power $\mathcal{P}_{lm}$ is cut-off for multipoles $l>2L$. Let us drop the factor of two and assume that $\mathcal{P}_{lm}=0$ for $l>L$, with $L$ even. In this case, $K(l_1, l_2; l, m)$ vanishes for $l>L$ and we can use the two-point function to directly determine the values of $K$  for $l=L$. At the end of this subsection, we shall comment on how to iteratively establish whether there exists a putative cut-off at $l=L$.

Consider equation (\ref{eq:two-point}) with $l_1=l$ and $l_2=l+L$. Then, due to the properties of angular momentum addition 
\begin{equation}\label{eq:two-point_atL}
	\langle a^*_{l m} a_{l+L,m+M}\rangle=(-i)^L
	D(l, m; L, M| l+L, m+M) K(l, l+L; L, M),
\end{equation}
which allows to directly solve for $K$ in terms of the two-point function. However, the two-point function is not directly observable. In order to measure  $K$ we need a sufficient number of realizations of the random process. At this point,  note that the  value of $K$ does not depend on  $m$. Hence we can define the following unbiased estimator,
\begin{equation}\label{eq:K_estimator}
	\hat{K}(l, l+L; L, M)\equiv \frac{ i^L}{2l+1}\sum_{m} \frac{a^*_{l m}a_{l+L, m+M}}{D(l, m; L, M | l+L, m+M)}.
\end{equation}
For $L=M=0$, our estimator $\hat{K}$  gives the $C_l$'s, equation (\ref{eq:C_l}). Therefore, $\hat{K}$ can be regarded as a generalization of the angular power spectrum. However, note that  for $L\neq 0$ the variable $\hat{K}$ is not a scalar, and, in particular, does not transform under an irreducible representation of the rotation group. Let us stress here an  additional property of $\hat{K}$. Namely, because $\hat{K}$ is  quadratic in the temperature fluctuations, our study of isotropy is decoupled of issues about non-Gaussianity.  If the expectation value of $\hat{K}$ does not vanish for $L\geq 0$, primordial perturbations are statistically anisotropic, irrespective of whether temperature anisotropies are Gaussian or not.   However, we cannot measure $\langle \hat{K}\rangle$ directly, and the different realizations of the random process $\hat{K}$ are expected to fluctuate around its expectation value. It is here where we have to make additional assumptions about the nature of the fluctuations. In order to estimate the  variance of $\hat{K}$, let us assume for simplicity that  perturbations are Gaussian and statistically isotropic. This assumption is justified because we are attempting to detect the existence of statistical anisotropies. Then, under these conditions, we find that the real and imaginary parts of $\hat{K}$ are uncorrelated, and for $L\neq 0$
\begin{equation}\label{eq:K_variance}
	\left\langle  
	(\hat{K}^*\hat{K})(l,l+L; L, M)\right\rangle
	=\frac{ C_l\, C_{l+L}}{(2l+1)^2}\sum_{m}
	\frac{1}{D(l, m; L, M | l+L, m+M)^2}.
\end{equation}  
We encounter here the same cosmic variance problem one faces when trying to measure the angular power spectrum. Because for a given $l$ we only have a finite number of samples of the two-point function, $2l+1$, we cannot measure $\langle\hat{K}\rangle$ with infinite precision. By the way, note that if  the value of the estimator happens to be  consistent with zero,  then we  guessed incorrectly, and primordial anisotropies are not cut-off at $l=L$, but eventually at $l=L-2$.

\subsection{A Different Estimator for $K$}

As we shall see, the orientation-dependence of our statistic $\hat{K}$  makes the analysis of temperature anisotropies rather cumbersome. Therefore, it may be advantageous to use an  estimator of $K$ that possesses  well-defined properties under rotations. In order to construct it, let us consider the quadratic combination
\begin{equation}\label{eq:B}
	B(l_1, l_2; l, m)=\sum_{m_1, m_2} a^*_{l_1 m_1} a_{l_2  m_2} (-1)^{m_1}
		\langle l_1,-m_1; l_2 m_2 | l, m\rangle,
\end{equation}
where $\langle l_1, m_1, l_2, m_2 | l, m\rangle$ is a Clebsch-Gordan coefficient. The expectation value of $B$ is nothing else but the bipolar spherical harmonic of Hajian and Souradeep \cite{HajianSouradeep}. It can be shown \cite{Varshalovich} that under rotations $B$ transforms under an irreducible representation of the rotation group,
\begin{equation}
	B(l_1, l_2; l, m)\to \sum_{\tilde{m}}D^{l}_{m\tilde{m}} \, B(l_1, l_2; l, \tilde{m})  ,
\end{equation} 
where $D$ is the Wigner matrix of the rotation.  Using equation (\ref{eq:two-point}), we can express the expectation value of $B$ (the bipolar coefficients) in terms of our quantity $K$,
\begin{equation}\label{eq:bipolar}
	\langle B\rangle =\sum_{m_1 m_2}\sum_{L M} (-i)^{l_2-l_1} D(l_1, m_1; L,M | l_2, m_2)
		K(l_1, l_2; L, M) (-1)^{m_1}\langle l_1, -m_1; l_2, m_2| l, m \rangle.
\end{equation}
Again, the previous expression simplifies if there is an anisotropy cut-off at multipoles $l>L$. In that case, one can use the previous expression as a guidance to derive an unbiased estimator of $K$,
\begin{equation}\label{eq:cal_K}
\hat{\mathcal{K}}(l, l+L; L,M)=\frac{i^L B(l, l+L; L, M)}{\sum_{m} (-1)^m D(l, m; L, M| l+L, m+M)
	\langle l,-m; l+L, m+M | L, M \rangle},
\end{equation}
which is guaranteed to transform appropriately under rotations. However, as opposed to the previous case, we only have one realization of the quantity we are trying to estimate, so our estimator suffers from a larger cosmic variance. As a matter of fact, the mean square fluctuations of $\hat{\mathcal{K}}$ are given by
\begin{equation}\label{eq:cal_K_variance}
	\left\langle (\hat{\mathcal{K}}^*\hat{\mathcal{K}})(l, l+L; L,M)\right\rangle=
	\frac{C_l C_{l+L}}
	{\sum_m  D(l, m ; L, M | l+L, m+M)^2},
\end{equation}
where we have used equations (\ref{eq:D}) and (\ref{eq:symmetry}) and, once again, we have assumed that perturbations are statistically anisotropic. It is apparent from the last equation, that the variance of $\hat{K}$, equation (\ref{eq:K_variance}), is smaller that the one of $\hat{\mathcal{K}}$. In this paper we shall concentrate on the lower variance estimator $\hat{K}$;  we reserve the study of $\hat{\mathcal{K}}$ and related  quantities to  a  hopefully forthcoming publication. 

To conclude this section, let us  comment on the relation between our integral $K$ and the bipolar spectrum we briefly mentioned during our previous discussion. In order to characterize the statistical anisotropy of the CMB sky, Hajian and Souradeep \cite{HajianSouradeep} define the bipolar spherical coefficient to be, up to a sign,  the expectation value of $B$, equation (\ref{eq:bipolar}). In that respect $K(l_1, l_2; l, m)$, equation (\ref{eq:two-point}),  and the bipolar spherical coefficients are similar: both vanish if the CMB sky is statistically isotropic and viceversa. To reduce the errors of their estimates, Hajian and Souradeep also define the bipolar spectrum, which is quadratic in the bipolar spectrum coefficients, and hence quartic in the $a_{lm}$'s.  We have not followed this path here, but have defined instead  suitable  estimators that are quadratic in the temperature fluctuations. The main difference between out approach and the one of \cite{HajianSouradeep} though is the physical focus. Whereas the  authors of \cite{HajianSouradeep} are concerned with the statistical properties of the CMB by itself, we are ultimately interested in the statistical properties of the primordial perturbations. Hence, our statistics $\hat{K}$  and $\hat{\mathcal{K}}$ (and as we shall see also $\hat{A}$ and $\hat{\mathcal{A}}$) mirror the properties of the primordial perturbations and owe its definition to a possible origin of statistical anisotropies in the CMB.
  
\subsection{Large Angular Scales}
Our analysis so far has been fairly general. We have not made any strong  assumptions about the nature of the primordial perturbations, the content of the universe or the angular scales involved. This is why our handle on statistical anisotropies is not as strong as it could be. At this point we shall trade generality for constraining power. On large angular scales, the integral $K$ can be explicitly evaluated, equation (\ref{eq:explicit_K}). Therefore, we can use large angular scales to explicitly determine the power in the primordial anisotropy multipoles. 

For simplicity, let us assume that the power $\mathcal{P}_{lm}$ is scale invariant in the scales of interest, $\mathcal{P}_{lm}\approx A_{lm}$. In a more sophisticated analysis we would try to fit for the value of the spectral index $n$. Setting accordingly $n=1$ in equation (\ref{eq:explicit_K}) we find that the integral is proportional to  
\begin{equation}
	I(l_1,l_2)=\frac{\pi}{72}\frac{\Gamma[2] \Gamma[(l_2+l_1)/2]}
		{\Gamma[(l_1-l_2+3)/2]\Gamma[(l_2-l_1+3)/2]\Gamma[(l_2+l_1+4)/2]},
\end{equation}
which does not depend on $l$ or $m$.   Therefore, substituting its value into equation (\ref{eq:K_estimator}) we arrive at the following estimator for the amplitude $A_{LM}$
\begin{equation}
	\hat{A}^{(l)}_{LM}\equiv \frac{i^L}{2l+1}
	\frac{1}{I(l,l+L)}
	\sum_{m} \frac{a^*_{lm}a_{l+L,m+M}}{D(l, m; L, M | l+L, m+M)},
\end{equation}
where $L$ is again the multipole above which the power in the multipoles is cut-off.  Note that we have actually found a set of estimators (one for each $l$) of the very same amplitude $A_{LM}$.  Hence, we can use this degeneracy to improve the errors of our measurements. Define the averaged estimator  
\begin{equation}\label{eq:A}
	\hat{A}_{LM}=\frac{1}{l_{max}-l_{min}+1}\sum_{l=l_{min}}^{l_{max}}
	\hat{A}^{(l)}_{LM}.
\end{equation}
In order to improve our measurement of $A_{LM}$, we would like the difference $l_{max}-l_{min}$ to be as large as possible. The values of $l_{min}$ and $l_{max}$ are essentially determined by the regime where our large-scale approximation is valid. Therefore, following our previous discussion, we will set $l_{min}=8$ and $l_{max}+L=30$ in future evaluations.   Assuming again that primordial perturbations are Gaussian and statistically isotropic we find that the mean square fluctuations of our estimator are
\begin{equation}
	\left\langle 
	|\hat{A}_{LM}|^2\right\rangle=
	\frac{1}{(l_{max}-l_{min}+1)^2}
	\sum_{l=l_{min}}^{l_{max}}
	\left\langle |\hat{A}^{(l)}_{LM}|^2\right\rangle,
\end{equation}
where the variances  of the individual estimators essentially follow from equation (\ref{eq:K_variance})
\begin{equation}
 	\left\langle 
	|\hat{A}^{(l)}_{LM}|^2\right\rangle=
	\frac{1}{I(l,l+L)^2}
 	\frac{C_{l} C_{l+L}}{(2l+1)^2}\sum_{m}
	\frac{1}{D(l, m; L, M | l+L m+M)^2}.
\end{equation}

So far we have concentrated on the amplitude of the primordial multipoles right below an eventually existing cut-off above $l=L$. As opposed to our previous discussion of the integrals $K$, in the case at hand it is possible to easily determine $A_{lm}$ for all values of $l\leq L$ by a recursive procedure.  Imagine we know $A_{\tilde{l}m}$ with sufficient accuracy for $\tilde{l}\geq l+1$. Then, using equation (\ref{eq:two-point}) we can solve for $A_{lm}$,
\begin{equation}
	A_{l m}=\frac{(-i)^{-l}
	\langle a^*_{l_1 m_1} a_{l_1+ l m+m_1}\rangle-
	\sum_{\tilde{l}= l+1}^{L} D(l_1, m_1; \tilde{l}, m | l_1+l, m_1+m) I(l_1, l_1+l) A_{\tilde{l}m}}
	{\,D(l_1, m_1; l, m | l_1+l, m_1+m)}.
\end{equation}
Because the left hand side of the equation does not depend on $m_1$, while the right hand side does, we could use it to construct an unbiased estimator for $A_{lm}$ and compute its variance, just as we did before. However, since we shall not need these estimators here, we shall not pursue this venue any further.

For completeness, let us  mention that in full analogy with the definition of $\hat{A}^l_{LM}$, one can also derive an estimator $\mathcal{A}$ of the amplitude $A_{LM}$ that transforms under an irreducible representation of the rotation group,
\begin{equation}
	\hat{\mathcal{A}}^l_{LM}=\frac{i^L}{I(l,l+L)}\frac{B(l, l+L ; L, M)}
	{\sum_m (-1)^m D(l,m; L, M | l+L, m+M)C(l,-m; L+L,M+M | L,M)}.
\end{equation}
Its variance  trivially follows from equation (\ref{eq:cal_K_variance}); as in the case of $\mathcal{K}$ the variance of $\hat{\mathcal{A}}$ is bigger than the one of $\hat{A}$. As mentioned earlier, we shall not further consider the estimator $\hat{\mathcal{A}}$ here.
 
\subsection{A preferred direction in the sky?}
\renewcommand\arraystretch{1}

Let us finally apply the formalism we have developed  to the anisotropies we observe in the CMB sky. Imagine that, for whatever reason, primordial perturbations display a preferred direction.  More precisely, suppose that primordial perturbations contain a dipolar component $\bar{\Phi}_{1m}$.  This dipole gives rise to a dipole $\mathcal{P}_{1M}$ and a quadrupole $\mathcal{P}_{2M}$ in the primordial spectrum. In the following we shall test for the existence of the quadrupole component. A discussion of the impact of a preferred direction on the  CMB sky can be also found in \cite{GoHuHuCr, MagueijoLand}.

Assume hence that perturbations in our universe are statistically anisotropic, and $\mathcal{P}_{lm}$ only vanishes for $l\geq 3$. We can then use the estimator defined in equation (\ref{eq:K}) to measure  $K(l, l+2; 2, M)$. If there is no power in the anisotropy multipole $L=2$, then ${K(l,l+L; 2, M)}$ vanishes. Therefore, the statistic $\hat{K}$ probes the existence of  preferred direction in the primordial perturbations. 

\begin{figure}
\includegraphics[width=8cm]{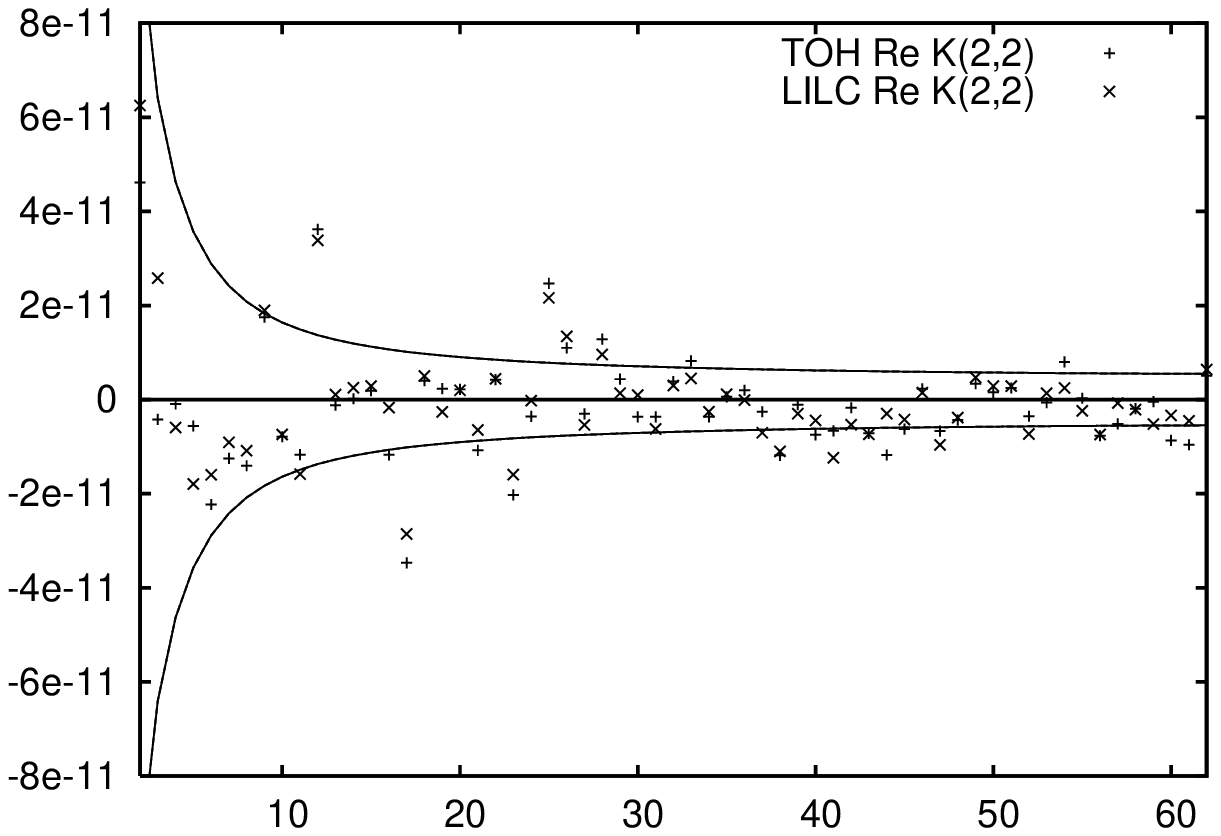}
\includegraphics[width=8cm]{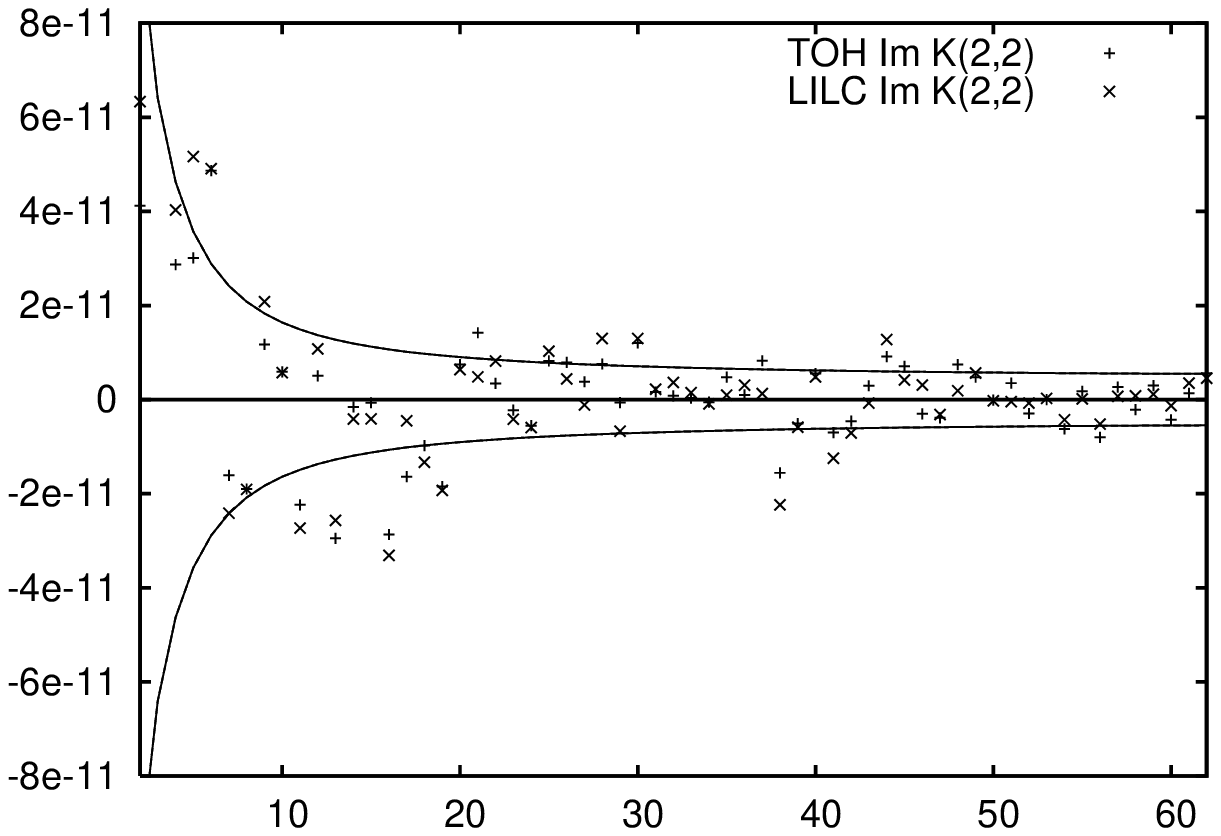}
\caption{Plots of the real and imaginary part of  $\hat{K}(l,l+2;2,2)$ (data points) and its root mean square fluctuation in a statistically isotropic universe (continuous  line).  \label{fig:K}}
\end{figure}

\begin{figure}
\includegraphics[width=8cm]{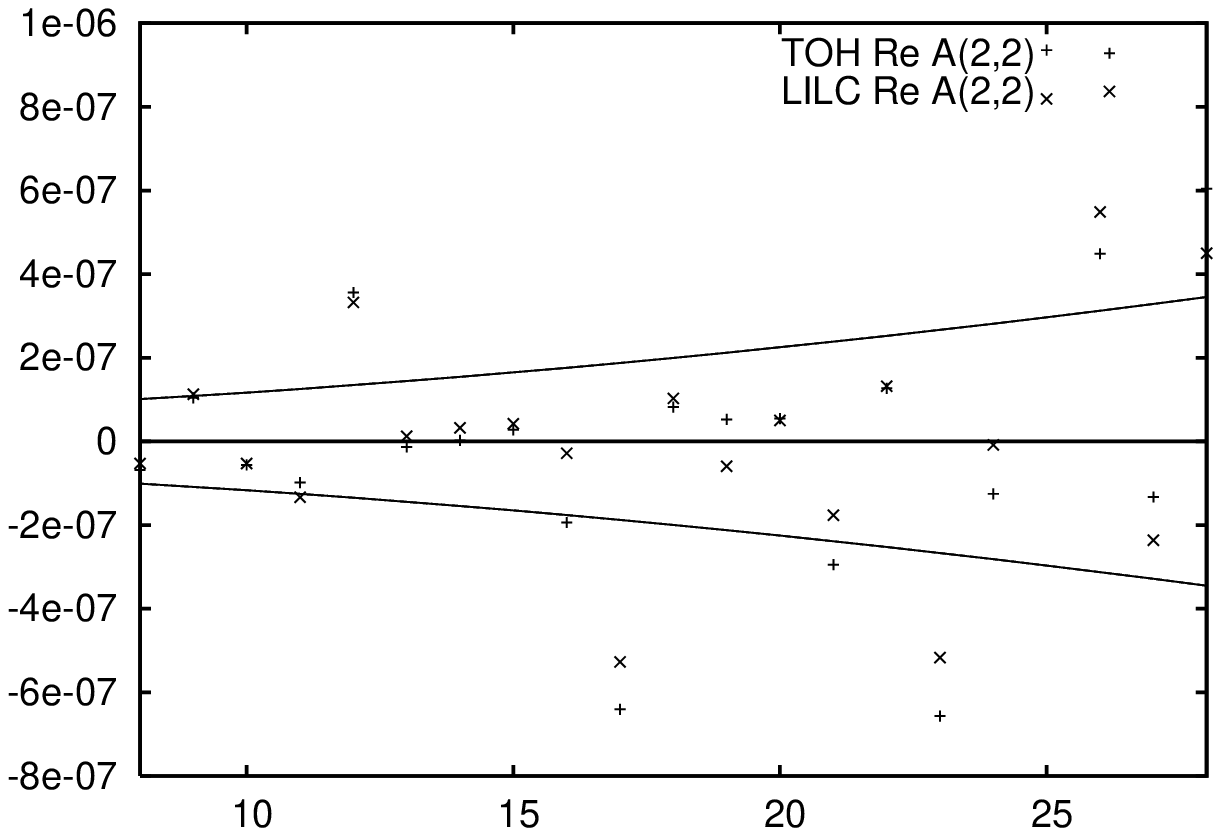}
\includegraphics[width=8cm]{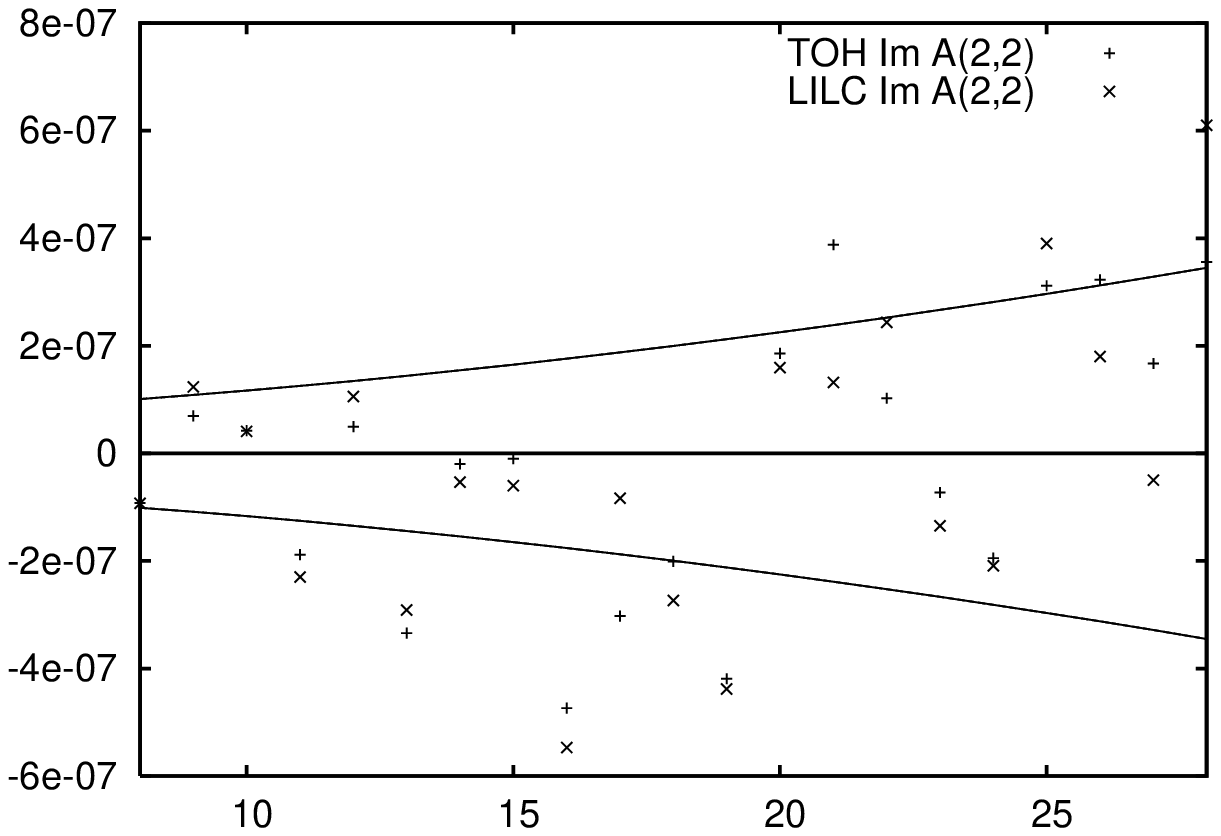}
\caption{Plots of the real and imaginary part of  $\hat{A}^{(l)}_{22}$ (data points) and its root mean square fluctuation in a statistically isotropic universe (continuous line).\label{fig:A}}
\end{figure}

We have numerically evaluated $\hat{K}(l,l+2; 2, M)$ using the foreground cleaned maps described in references \cite{TegmarkOliveiraHamilton} (henceforth TOH) and  \cite{LILC} (henceforth LILC). Any conclusion derived from these maps should be interpreted with care, as it has been argued that they are contaminated by  residual foregrounds  and their noise properties are ill-understood \cite{LILC}.  However, because sky cuts in CMB maps introduce artificial anisotropies \cite{HajianSouradeep}, at this point we are forced to proceed with these all-sky maps, in the hope that they accurately capture the real temperature fluctuations \cite{multipole-vectors}. 

For illustration, in Figure \ref{fig:K} we plot the real and imaginary parts of $\hat{K}(l,l+2; 2, M)$ for values of $l$ ranging between $l=2$ and $l=62$  and $M=2$.  Our choice of the upper $l$ limit is dictated by our neglect of instrumental noise in the error budget. Indeed, the errors in WMAP's  measurement of the power spectrum \cite{WMAP} are cosmic variance dominated (they contribute more than $90\%$ of the total error) only up to a multipole of about $l=64$. The error contours delimit the mean square fluctuations of the estimator under the assumption of isotropy, equation (\ref{eq:K_variance}). In order to evaluate the latter quantities, we use the $\Lambda$CDM model that best fits the WMAP data only \cite{WMAP}.

\begin{figure}
\includegraphics[width=8cm]{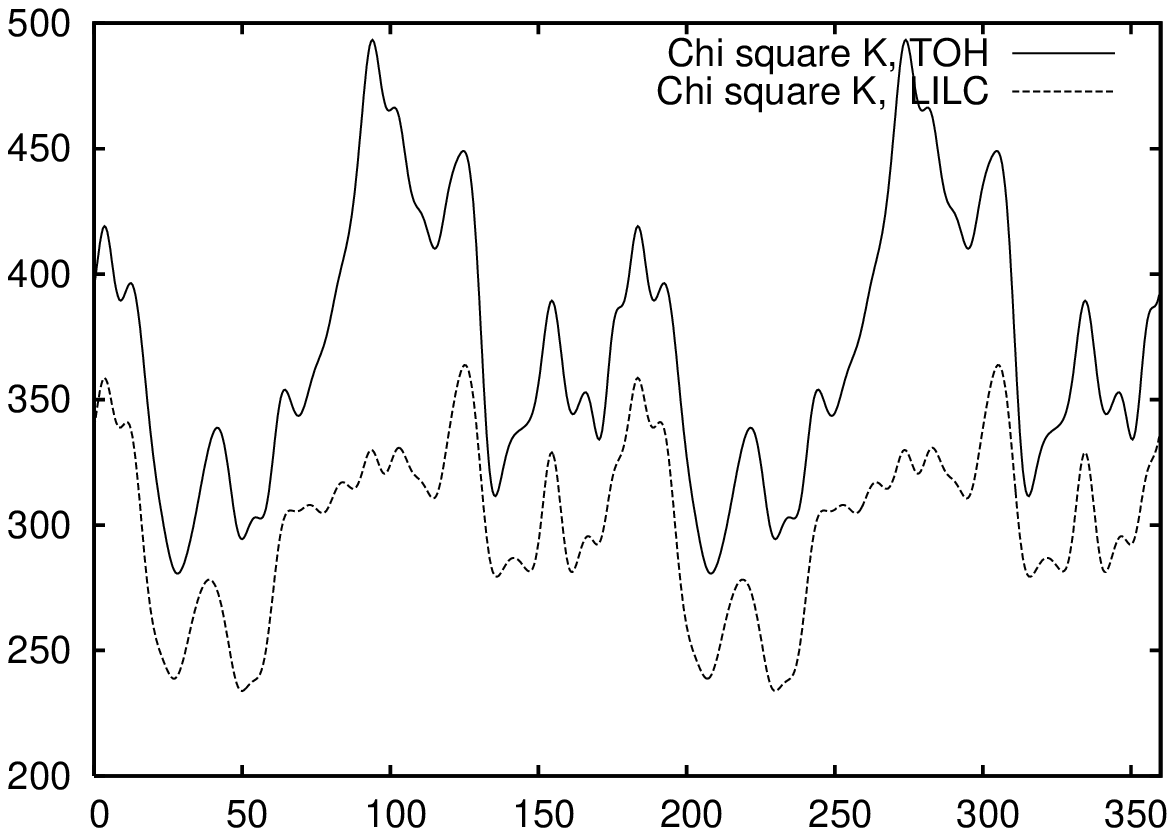}
\includegraphics[width=8cm]{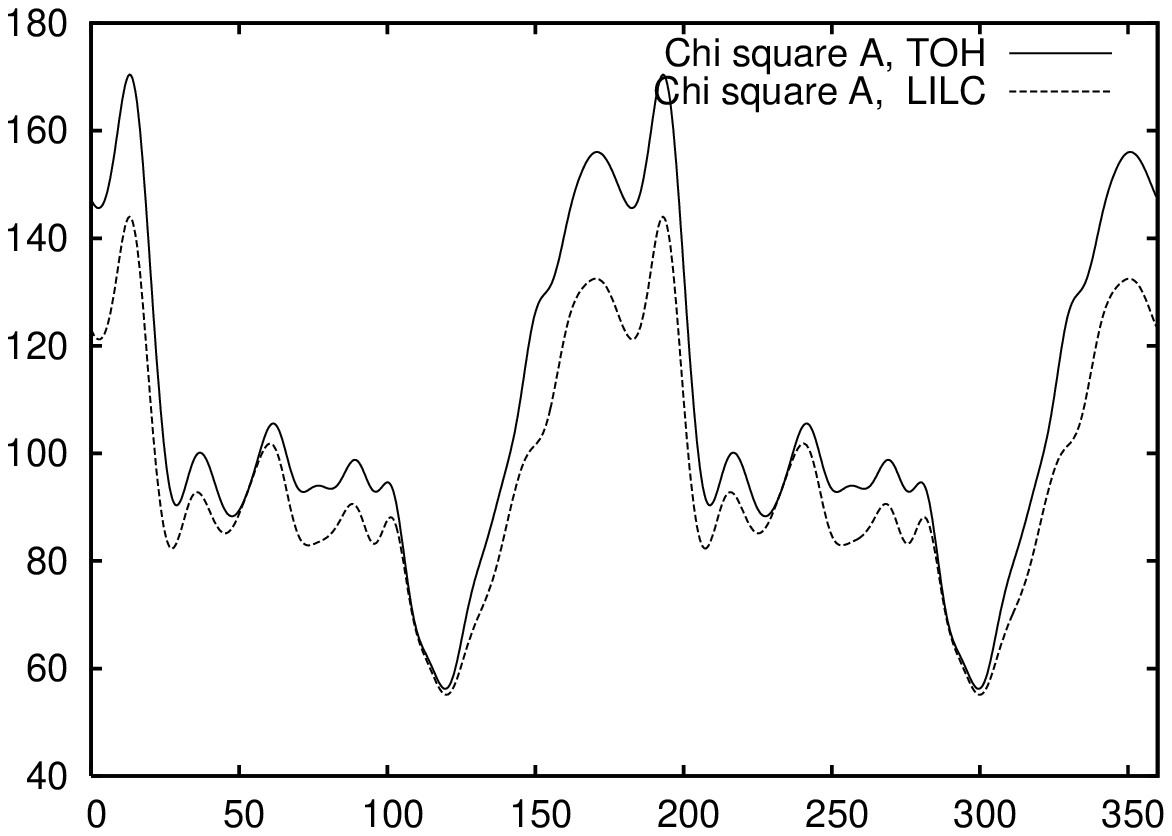}
\caption{The overall $\chi^2$ as a function of rotation angle. The CMB map is rotated along the $x$-axis (left) and $y$-axis (right) of the galactic coordinate system. Note that the plots have a period of $180^\circ$. Chi square is invariant under rotations around the $z$-axis (not shown).   \label{fig:orientation}}
\end{figure}

It is evident from the figure that both TOH and LILC maps lead to similar but not quite identical results. We can  determine whether data is well-fit by statistical isotropic perturbations by computing the value of $\chi^2$ for the statistic $\hat{K}$.  In order to determine how $\chi^2$ is distributed,  we have generated $64\cdot 10^4$ random skies drawn from a statistically isotropic distribution with WMAP's best fit angular power spectrum. The fraction of realizations $P_\textrm{rdm}(\chi^2<\chi_0^2)$ that have an overall $\chi^2$ smaller than the  one of the CMB sky in galactic coordinates is $97\%$ for the TOH map and $79\%$ for  the LILC map. However, these results are misleading, as  it turns out that the values of $\chi^2$ strongly depend on the orientation of the map. This is due to the non scalar nature of 
\begin{equation}
	\chi^2=\sum_l \sum_{M=0}^{L} \left(1+\delta_{M 0}\right)
	\frac{|\hat{K}(l, l+2; 2, M)|^2}{\langle |\hat{K}(l, l+2; 2, M)|^2\rangle},
\end{equation}
which is a consequence of the peculiar transformation properties of our estimator under rotations. In Figure \ref{fig:orientation} we plot how $\chi^2_0$ changes as a function of the map orientation. As seen in the figure, the value of $\chi^2_0$ widely varies with the orientation, so it is unclear at this point whether the map is consistent with statistical isotropy.

\begin{figure}
\includegraphics[width=8cm]{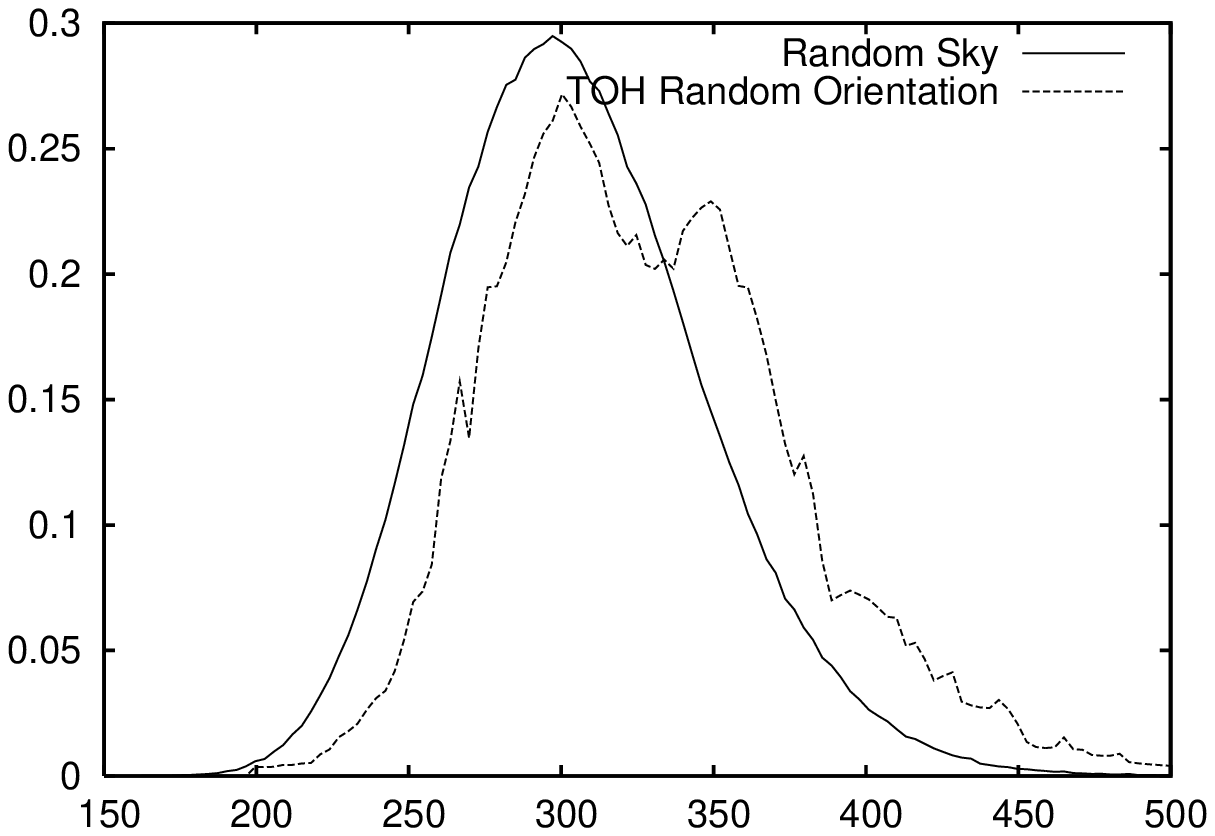}
\includegraphics[width=8cm]{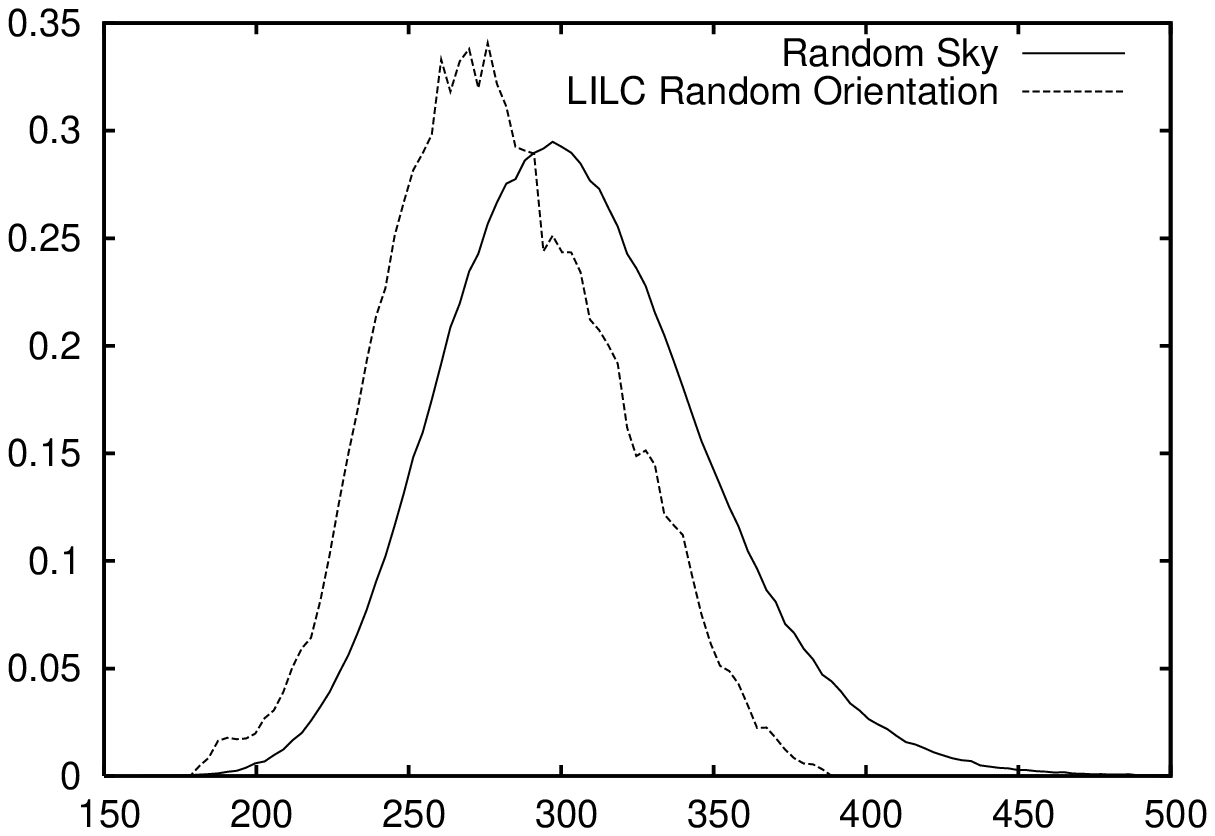}
\caption{Comparison of the distribution of $\chi^2$ values of $\hat{K}$ for statistically isotropic random skies and random orientations of the actual CMB map. The distributions have been binned to reduce noise.\label{fig:comparison}}
\end{figure}

If the CMB sky is statistically isotropic, as we rotate the temperature anisotropy map into all possible orientations, we expect the corresponding values of $\chi^2_0$ to sample the distribution of random skies $P_\textrm{rdm}(\chi_0^2)$ we have considered earlier. In Figure \ref{fig:comparison} we compare the probability distribution function of $\chi^2$ for an isotropic sky with the distribution $P_\textrm{map}(\chi^2_0)$ we obtain by randomly orienting the actual CMB map. At first sight,  it seems  that both distributions are quite similar. In order to quantify the  significance of their agreement or eventual disagreement, let us consider the probability $P$ that the $\chi^2$ of a statistically isotropic sky is less than the one of the randomly oriented actual CMB map,
\begin{equation}\label{eq:P}
	P=\int d\chi^2_0 \, P_\textrm{rdm}(\chi^2<\chi^2_0)\cdot P_\textrm{map}(\chi^2_0),
\end{equation}
where, again,  $P_\textrm{rdm}(\chi^2<\chi^2_0)$ is the fraction of statistically isotropic  random skies with a chi square less than $\chi_0^2$, and $P_\textrm{map}(\chi^2_0)\, d\chi_0^2$ is the probability that the randomly oriented actual CMB mas has chi square equal to $\chi^2_0$. 

If the chi square values of a map happened to be invariant under rotations and equaled $\chi_0^2$, we would have $P_\textrm{map}(\chi^2)=\delta(\chi^2-\chi_0^2)$, so  $P$ would simply be $P_\textrm{rdm}(\chi^2<\chi_0^2)$. On the other hand, if both $P_\textrm{rdm}$ and $P_\textrm{map}$ represented the same distribution we would get $P=50\%$. Integrating equation (\ref{eq:P}) numerically we find
\begin{equation}
	P=64\% \quad \text{(TOH)} \quad \text{and} \quad P= 33\% \quad \text{(LILC)},
\end{equation}
that is, there is no significant evidence for statistical anisotropy.
  
We can try to strengthen these results by considering large angular scales. In Figure \ref{fig:A} we plot the values of $\hat{A}^{(l)}_{2, M}$ for $M=2$ and $l_{min}=8<l<l_{max}=28$.   Again, as shown in Figure \ref{fig:orientation}, the values of $\chi^2$  depend on the map orientation. In order to quantify up to what extent  the maps are statistically isotropic on large angular scales, we have computed the probability $P$ in equation (\ref{eq:P}), applied in this case to the chi square of $\hat{A}^l_{LM}$. 
The probabilities are
\begin{equation}
	P=51\% \quad \text{(TOH)} \quad \text{and} \quad P= 35\% \quad \text{(LILC)},
\end{equation}
which agree quite well with the previous values. This is to be expected to some extent, since $\hat{K}$ and $\hat{A}$ are proportional to each other. However, note that the $\chi^2$ of $\hat{K}$ and $\hat{A}$ involve sums over different ranges of $l$. 

\begin{figure}
\includegraphics[width=8cm]{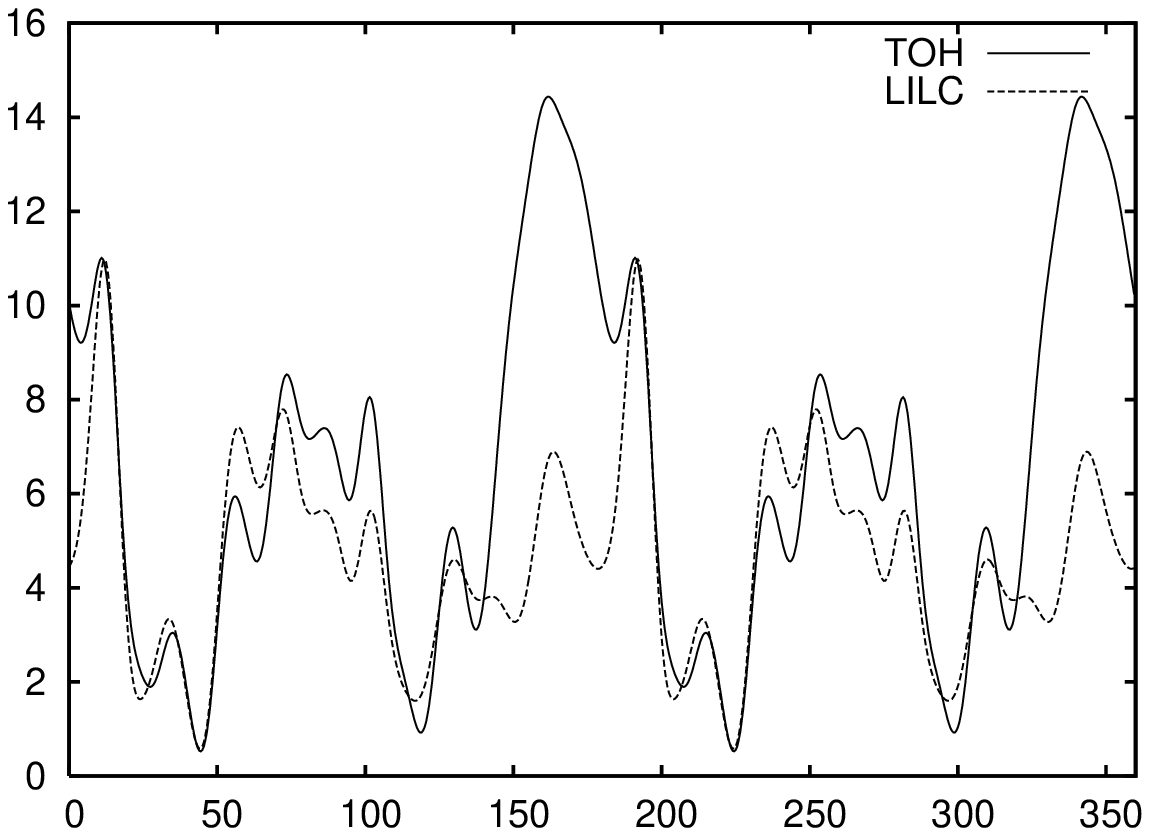}
\includegraphics[width=8cm]{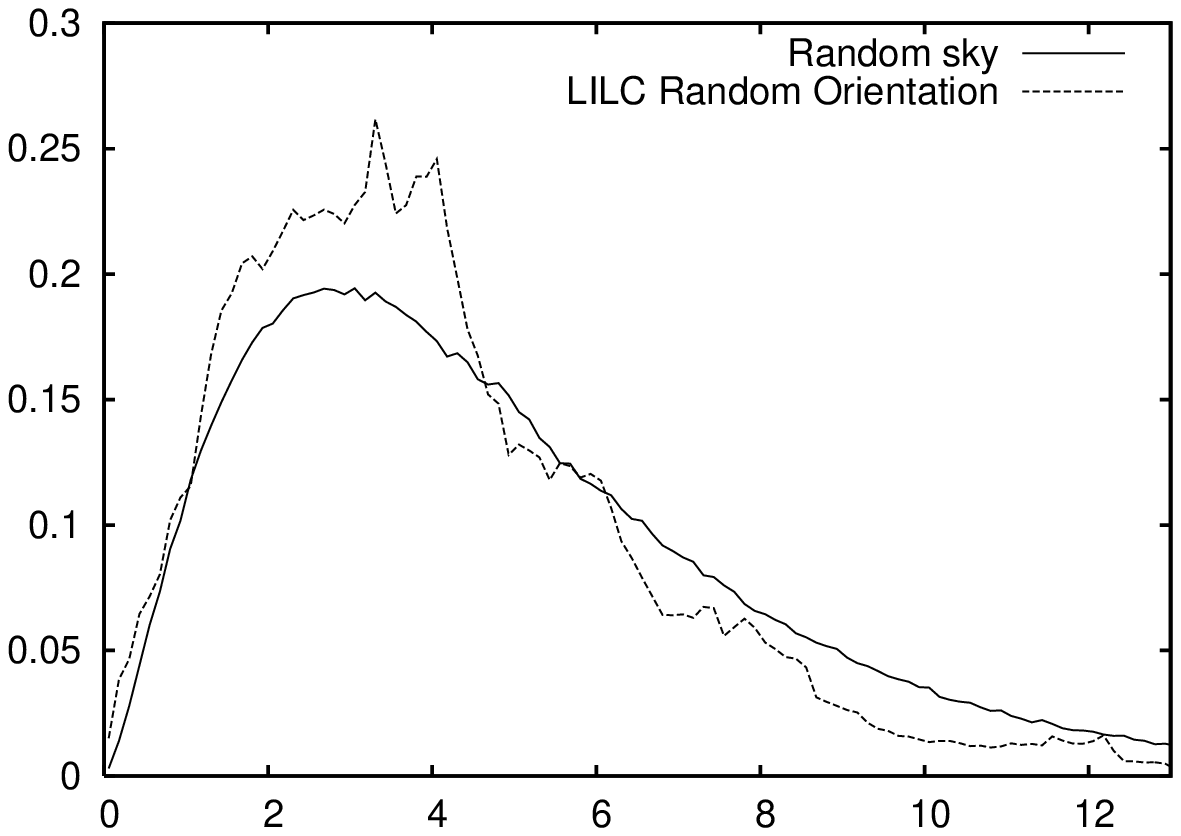}
\caption{In the left panel we show the overall $\chi^2$ as a function of rotation angle. The CMB map is rotated along the $y$-axis in galactic coordinates. In the right panel we plot the distribution of values of $\chi^2$ for randomly generated skies and randomly chosen orientations of the LILC map.   \label{fig:A_tot}}
\end{figure}

Finally, to confirm our conclusions, let us evaluate the estimator for the amplitude of an eventual quadrupole anisotropy component, equation (\ref{eq:A}). In Figure \ref{fig:A_tot} we plot how the corresponding $\chi^2$ varies as a function of the map orientation and compare, as in the previous cases,  randomly simulated skies with random orientations of the actual CMB sky.  In this case, signs of statistical anisotropies  are still absent. Indeed,  the   probability (\ref{eq:P}) for the statistic $\hat{A}_{LM}$ is
\begin{equation}
	P=48\% \quad \text{(TOH)} \quad \text{and} \quad P= 43\% \quad \text{(LILC)}.
\end{equation}

\section{Summary and Conclusions} \label{sec:conclusions}

In this paper we have proposed a model-independent parametrization of the primordial spectrum that accommodates statistical anisotropies.  We have determined how these anisotropies impact the temperature fluctuations in the cosmic microwave background, and what type of footprints  are left behind by the former. Remarkably, the angular power spectrum does not differentiate between statistically isotropic and anisotropic perturbations. As far as the $C_l$'s  are concerned, the angular power spectrum is as well fit by an isotropic primordial spectrum than by an anisotropic one. 

In order to find out whether primordial perturbations are statistically isotropic, one has to study the non-diagonal components of the two-point correlation function. In this paper, we have considered a set of statistics  that measure the amplitude of the different multipoles in the primordial perturbations. The first set, $\hat{K}$ and $\hat{A}$, does not transform under an irreducible representation of the rotation group, but has a lower variance than the second set, $\hat{\mathcal{K}}$ and $\hat{\mathcal{A}}$, which does possess well-defined transformation properties under rotations. In this paper, we have concentrated on the first set, leaving the second for future work.  When applied to the full-sky maps of the CMB temperature fluctuations currently available, the former statistics do not lead to evidence for statistical anisotropy, although the analysis is obscured by the strong variations of our estimators with the  map orientation.  

The statistical isotropy of the primordial perturbations is an observational issue. As with Gaussianity, scale invariance or adiabaticity, it should be constrained rather than postulated or automatically implied. In this paper we have taken the first steps in that direction. If at the end of the journey it turns out that  observational data is  inconsistent with isotropic primordial perturbations, our current (inflationary) models for  the origin of structure will have to be discarded and replaced by radically different ones. 

\begin{acknowledgements}
Some of the results in this paper have been derived using the GNU GSL numerical libraries and the HEALPix\footnote{\texttt{http://healpix.jpl.nasa.gov/}.}
\cite{GorskiHivonWandelt} and \nobreak{CMBEASY} \cite{CMBEASY} software packages. The author thanks Simon Catterall, Wayne Hu, Dragan Huterer, Levon Pogosian and Mark Trodden for valuable feedback and comments, and Benjamin Wandelt for useful conversations during the embrionary stages of this project. I am particularly grateful to  Wayne Hu for  inquiring about the behavior of our estimators under rotations.
\end{acknowledgements}

\appendix

\section{Useful Formulae}
In this appendix we list a series of useful formulae, mainly concerning spherical harmonics. We follow the conventions in \cite{Sakurai} (which agree with the ones of \cite{GorskiHivonWandelt}.)

\begin{itemize}

\item Expansion of plane wave in Legendre polynomials
\begin{equation}
	\exp(i \vec{k}\cdot\hat{n})=\sum_l (2l+1)\, i^l\, j_l(k)\, P_l(\hat{k}\cdot\hat{n}).
\end{equation}

\item Addition theorem
\begin{equation}\label{eq:addition-theorem}
P_l(\hat{k}\cdot \hat{n})=\frac{4\pi}{2l+1}\sum_m Y_{lm}(\hat{k}) Y^*_{lm}(\hat{n}).
\end{equation}

\item Orthonormality of spherical harmonics
\begin{equation}\label{eq:orthonormality}
	\int d^2{\hat{k}} \,\, Y^*_{lm}(\hat{k}) Y_{l'm'}(\hat{k})
	=\delta_{ll'}\delta_{mm'}.
\end{equation}
\item Behavior under complex conjugation
\begin{equation}\label{eq:complex-conjugation}
 Y^*_{lm}=(-1)^m Y_{l-m}.
\end{equation}
\item Addition of angular momenta
\begin{equation}
	Y_{l_1m_1}(\hat{k}) Y_{l_2 m_2}(\hat{k})=\frac{1}{\sqrt{4\pi}} \sum_{lm}
	D(l_1, m_1; l_2, m_2| l, m)\, Y_{lm}(\hat{k}),
\end{equation}
where 
\begin{equation}\label{eq:D}
	D(l_1, m_1; l_2, m_2| l, m)=\sqrt{\frac{(2l_1+1)(2l_2+1)}{(2l+1)}}
	\langle l_1, 0; l_2, 0 | l, 0\rangle
	\langle l_1, m_1; l_2, m_2 | l, m\rangle.
\end{equation}
The $\langle l_1, m_1; l_2, m_2 | l, m\rangle$ are (real) Clebsch-Gordan coefficients.

\item Symmetry properties of Clebsch-Gordan coefficients
\begin{equation}\label{eq:symmetry}
	\langle l_1, m_1; l_2, m_2 |  l, m\rangle=
	(-1)^{l_1-m_1} \sqrt{\frac{2l+1}{2l_2+1}}\langle l_1, -m_1; l, m |  l_2, m_2\rangle.
\end{equation}

\item Integral of three spherical harmonics
\begin{equation}\label{eq:three_spherical}
	\int d^2\hat{k}\, Y_{lm}^*(\hat{k})
	Y_{l_1 m_1}(\hat{k})  Y_{l_2 m_2}(\hat{k})
	=\frac{1}{\sqrt{4\pi}}D(l_1, m_1; l_2, m_2| l, m).
\end{equation}
Note that $\sum_m Y^*_{lm} Y_{lm}$ is invariant under spatial rotations. Hence, it follows from equations (\ref{eq:three_spherical}) and (\ref{eq:orthonormality}) that 
\begin{equation}\label{eq:D_identity}
	\sum_m D(l, m; l_2, m_2 | l, m)=(2l+1)\, \delta_{l_2 0} \, \delta_{m_2 0}.
\end{equation} 

\item Relation between Clebsch-Gordan coefficients and Wigner's 3-j symbol
\begin{equation}
	\langle l_1, m_1; l_2, m_2| l_1, l_2; l, m\rangle
	= (-1)^{l1-l2+m} \sqrt{2l+1} 
	\left(
	\begin{array}{ccc} l_1 & l_2 & l \\ m_1 & m_2 & {}-m \end{array}
	\right).
\end{equation}

\end{itemize}

\end{document}